\documentstyle[psfig,aaspp4,11pt,multicol]{article}

\lefthead{J.~M.~Miller et al.}  \righthead{High Frequency
Quasi-periodic Oscillations in the 2000 Outburst of the Galactic
Microquasar XTE~J1550-564}

\received{Date}
\revised{Date}
\accepted{Date}

\journalid{vol}{date}
\articleid{1}{4}
\paperid{id}

\cpright{AAS}{1999}
\ccc{x}

\newcommand{\xtej}{\mbox{XTE~J1550$-$564}}

\begin{document}

\title{High Frequency Quasi-periodic Oscillations in the 2000 Outburst
of the Galactic Microquasar \xtej}

\author{J.~M.~Miller\altaffilmark{1},
        R.~Wijnands\altaffilmark{1,6},
        J.~Homan\altaffilmark{2},
	T.~Belloni\altaffilmark{3},
	D.~Pooley\altaffilmark{1},
	S.~Corbel\altaffilmark{4},\\
	C.~Kouveliotou\altaffilmark{5},
	M. van der Klis\altaffilmark{2}, 
	W.~H.~G.~Lewin\altaffilmark{1}
	}

\altaffiltext{1}{Center~for~Space~Research and Department~of~Physics,
        Massachusetts~Institute~of~Technology, Cambridge, MA
        02139--4307; jmm@space.mit.edu, rudy@space.mit.edu,
        davep@space.mit.edu, lewin@space.mit.edu}
\altaffiltext{2}{Astronomical Institute 'Anton Pannekoek,' University
        of Amsterdam, and Center for High Energy Astrophysics,
        Kruislaan 403, 1098 SJ, Amsterdam, the Netherlands,
        homan@astro.uva.nl,michiel@astro.uva.nl}
\altaffiltext{3}{OAB Brera, Italy; belloni@merate.mi.astro.it}
\altaffiltext{4}{Univsersit\'e Paris VII and Service d'Astrophysique,
        CEA Saclay, 91191 Gif sur Yvette, France,
        corbel@discovery.saclay.cea.fr} 
\altaffiltext{5}{NASA/MSFC, SD-50, 320 Sparkman Drive, Huntsville, AL,
        35812; chryssa.kouveliotou@msfc.nasa.gov}
\altaffiltext{6}{\it Chandra fellow}


\authoremail{jmm@space.mit.edu}

\label{firstpage}

\begin{abstract}
We present an analysis of the high frequency timing properties of the
April-May 2000 outburst of the black hole candidate and Galactic
microquasar \xtej, measured with the \textit{Rossi X-ray Timing
Explorer}.  The rapid X-ray variability we measure is consistent with
the source being in either the ``very high'' or ``intermediate'' black
hole state.  Strong (5--8\% rms) quasi-periodic oscillations (QPOs)
are found between 249-276~Hz; this represents the first detection of
the same high frequency QPO in subsequent outbursts of a transient
black hole candidate.  We also present evidence for lower-frequency
QPOs at approximately 188~Hz (3.5$\sigma$, single trial), also
reported previously and likely present simultaneously with the
higher-frequency QPOs.  We discuss these findings within the context
of the 1998 outburst of \xtej, and comment on implications for models
of QPOs, accretion flows, and black hole spin.
\end{abstract}


\section{Introduction}
The sub-class of X-ray binaries identified as ``microquasars'' is
growing steadily as efforts increase to observe both persistent and
transient systems simultaneously in the X-ray and radio bands.  Yet,
only two transient black hole candidate (BHC) systems with apparently
superluminal jet motion have been observed in repeated outbursts
(separated by clear quiescent periods) reaching flux levels in excess
of 0.5~Crab in soft X-rays: GRO~J1655$-$40 and \xtej.  Similar jets
and luminosities have been observed by many in GRS~1915$+$105; this
source has been in outburst continuously during the RXTE era.

The X-ray lightcurves of BHCs are usually described in terms of
canonical states characterized by specific spectral and timing
properties.  The details of these states are discussed at length
elsewhere (for reviews, see Tanaka \& Lewin 1995, and van der Klis
1995; for a recent discussion see Homan et al. 2001).  We merely note
the states briefly: the Very High State (VHS) is often most luminous,
may contain significant flux from both soft ($<$5~keV; usually a
multi-color disk blackbody as per Mitsuda et al.\ 1984) and hard
($>$5~keV; usually a power-law) flux components, and quasi-periodic
oscillations (QPOs) and/or significant timing noise; the High State
(HS) is strongly dominated by the soft flux and has weak ($\sim$few
per cent rms) power-law-like timing noise; the Intermediate State (IS)
is very similar to the VHS but less luminous; the Low State (LS) is
dominated by the hard component, has strong band-limited timing noise
($>$20\% rms), and sometimes QPOs; and the Quiescent or ``Off'' State
(QS), which is characterized by a very weak spectral power-law.

The 1998--1999 outburst of \xtej~(Smith 1998) was one of the most
remarkable yet observed from a BHC system.  A quasi-periodic
oscillation (QPO) with a frequency as high as 285~Hz was observed
(Remillard et al. 1999a; Homan, Wijnands, \& van der Klis 1999; Homan
et al. 2001), X-ray state transitions were observed which challenge
the primacy of the mass accretion rate $\dot{m}$ in driving outbursts
(Homan et al. 2001), and radio monitoring revealed jet production with
apparent velocity of v$_{\rm jet} > 2c$ (Hannikainen et al. 2001).

After a quiescent period of several months, on 2 April 2000, new
activity in \xtej~was noticed with the \textit{RXTE} All Sky Monitor
(ASM, Levine et al. 1996; \xtej~Smith et al.\ 2000), and the ensuing
outburst lasted more than 70 days.  This outburst reached a peak flux
of $\sim$1~Crab in the ASM (1.5--12~keV, see Figure 1).  The source
was seen out to 300~keV with BATSE aboard \textit{CGRO}, and
simultaneously in optical bands (Masetti \& Soria 2000; Jain, Bailyn,
Orosz, et al.\ 2000).  Evidence for emission from a compact jet (an
inverted radio spectrum) was reported during the
hard-X-ray-flux-dominated late decay phase of this outburst; a
possible discrete ejection event may have been observed during the
rapidly rising phase (Corbel et al. 2001).

On 28 January, 2001, activity was noticed in \xtej~for a third time.
\textit{RXTE} found the source to be in the LS, with timing noise of
40\% rms and an energy spectrum characterized by a power-law with
photon index $\Gamma=1.52$ (Tomsick et al. 2001).

The second outburst met the trigger criteria for our \textit{RXTE}
Cycle 5 Target of Opportunity program to study a BHC in outburst
(\textit{Chandra} spectroscopy of this outburst is detailed in Miller
et al. 2001b, and radio results in Corbel et al. 2001).  We observed
\xtej~with \textit{RXTE} on 18 occasions totaling 54.9~ksec of data
between 24 April and 12 May 2000.  We report QPOs with a frequencies
ranging between 249--278~Hz.  This finding is discussed within the
context of the previous outburst of \xtej, the behavior observed in
BHC/microquasars GRO~J1655$-$40 and GRS~1915$+$105, and models for the
inner accretion environments in these and other BHC systems.

\section{Observations}
In stark contrast to the 1998--1999 outburst of \xtej, the 2000
outburst lightcurve roughly follows a fast-rise exponential-decay (or,
``FRED'') profile.  Based on the (5--12~keV)/(3--5~keV) hardness ratio
accompanying the 1.5--12~keV ASM lightcurve in Figure 1, the spectral
analysis presented in Miller et al. (2001b), and the timing results we
detail in this work, we identify three outburst states: an initial LS,
a VHS/IS, and a final LS.  Whereas we can identify the middle state
using timing, spectral, and hardness signatures, the initial and final
state identifications are less certain.  We have only one PCA
observation in the initial LS, the latter LS identification is based
only on the ASM hardness ratio as we have no PCA observations within
this state (for timing studies of the LS, see Kalemci et al. 2001).
The middle state is not a HS episode: the soft component does not
strongly dominate the X-ray flux and power-law timing noise is not
observed.  This state is the most luminous within the 2000 outburst,
but it is less luminous than the VHS reported in the 1998--1999
outburst (fast QPOs were observed in both the VHS and IS during the
1998--1999 outburst).  Although this state is more likely an IS
episode, as a clear determination cannot be made we refer to the
middle state simply as the VHS/IS.

In all observations, data were collected in the Standard~1 (1/8 second
time resolution in one photon energy channel for the energy range
2--60~keV) and the Standard2f (129 channels for 2--60~keV, 16~seconds
time resolution) modes.  Simultaneously, data were collected in
several single bit and event modes.  In the first three observations
wherein we report a high frequency QPO (see Table 1), these modes are
SB\_125us\_0\_17\_1s (128$\mu$s time resolution in one 2--7.5~keV
energy channel), SB\_125us\_18\_35\_1s (128$\mu$s time resolution in
one 7.5--15~keV energy channel), and E\_16us\_16B\_36\_1s (16$\mu$s
time resolution in 16 channels from 15--60~keV).  In the latter five
observations, these modes are SB\_125us\_0\_13\_1s (128$\mu$s time
res. in one 2--6~keV energy channel), SB\_125us\_14\_35\_1s (128$\mu$s
time resolution in one 6--15~keV energy channel), and
E\_8us\_32M\_36\_1s (8$\mu$s time resolution in 32 channels from
15--60~keV).  Due to the fact that the modes in the early observations
do not cover exactly the same channels as the modes used in the latter
observations, we cannot examine the energy dependence of the fast QPO
in exactly the same energy ranges across the outburst, although the
differences are very small.

\section{Analysis and Results}
We made power spectra, using 16~second data segments, of the combined
single-bit and event modes data.  In order to explore the
high frequency regime in this outburst, we analyze the range
64--4096~Hz.  Although we detect QPOs between 1--10~Hz similar to
those reported in the first outburst of \xtej~(e.g., Homan et al.\
2001), the analysis and interpretation of these features is
complicated and left to future work.

We fitted the 64--4096~Hz range with a model consisting only of a
Poisson noise component and a single Lorentzian; we obtain reduced
$\chi^{2}$ values roughly between 1--1.3 using this model (for 176
degrees of freedom).  We report QPOs with frequencies varying between
249--278~Hz, significant at 4.4--6.2$\sigma$, exclusively in
observations during the VHS/IS, though not in all VHS/IS observations.
The significance of the QPOs is highest at the start of the VHS/IS and
decreases steadily (though not monotonically) through the state.  We
report the QPO parameters in Table 1 in the energy band where the
features are most significant (6--60~keV) for observations where a QPO
is clearly present; the quoted errors are for $\Delta\chi^{2}~=~1$
(1$\sigma$).

The strength of the QPOs is also examined in the 2--60~keV, 2--6~keV,
6--15.3~keV, and 15.3--60~keV bands (bands chosen as single-bit and
event mode binning allowed).  As is typical for high frequency QPOs in
BHCs, the QPOs are stronger in the higher energy bands and weaker when
the 2--6~keV band is included.  In Figure 2, we show how the QPO
parameters vary as a function of time and as a function of 2--60~keV
flux.  The strength decreases with time, and increases with flux;
these findings are consistent with the QPO significance decreasing as
the VHS/IS flux decays.  The QPO FWHM is constant.  Within errors, the
QPO frequency decreases monotonically with time, an increases
monotonically with 2--60~keV flux.

We also analyzed the strength, FWHM, and frequency of the QPO versus
three hardness ratios: (6--15~keV)/(2--6~keV),
(15--60~keV)/(6--15~keV), and (15--60~keV)/(2--6~keV) (see Figure 3).
The frequencies and strengths are weakly anti-correlated with
increasing (6-15~keV)/(2--6~keV) hardness, weakly positively
correlated with increasing (15--60~keV)/(6--15~keV) hardness, and
weakly anti-correlated with increasing (15--60~keV)/(2--6~keV)
hardness.

We measure upper limits on the strength of possible QPO features on
days where the 268~Hz QPO is not clearly observed.  For observations
made on days (of the year 2000) 114.6, 116.1, 118.2, 123.7, we report
95\% confidence upper limits for the 268~Hz QPO between 2.7--5.2\% rms
for a feature with FWHM fixed at 50~Hz in the 64--512~Hz range
(6--60~keV).  We note that the upper limit obtained for the only
observation for which \xtej~is in the LS (day 114.6; see Figure 1) is
2.7\%.  For observations made on days 125.2, 126.8, 127.4, 128.4,
128.9, 130.8, 131.6, and 132.4, the 95\% confidence upper limits range
between 5.1--6.4\% rms.  Any fast QPO in these observations is
significant at only the 2$\sigma$ level, or less.  Thus, the 268~Hz
QPO is only clearly detected in the observations which fall within the
VHS/IS.  At the time of writing, observations made during the LS
episode following the VHS/IS are not yet public, however Kalemci et
al. (2001) report on timing studies of the LS and find no evidence for
the 268~Hz QPO.

We note evidence for a second fast QPO, at $188\pm3$~Hz.  In simply
adding all six observations in which we report the presence of the
higher frequency QPO (see Figure 4), we find that the lower QPO is
significant at 3.5$\sigma$ ($24^{+33}_{-9}$~Hz FWHM,
$2.8^{+0.9}_{-0.4}$\% rms) and the higher QPO is significant at
7.8$\sigma$ ($56^{+8}_{-7}$~Hz FWHM, $6.2^{+0.4}_{-0.4}$\% rms).
Adding all six observations with the higher frequency QPO gives 22
trials for the 268~Hz QPO and 48 trials for the 188~Hz QPO (quoted
significances are single-trial).  For convenience, we now refer to the
higher-frequency feature as the 268~Hz QPO, and to the lower-frequency
feature as the 188~Hz QPO.  Like the 268~Hz QPO, the 188~Hz QPO is
also strongest in the 6--60~keV band.

We investigated how the strength of the QPOs as seen in the summed
data (see Figure 4) depends on energy in the 2--6~keV, 6--15.3~keV,
and 15.3--60~keV bands (corresponding to the bands in Table 1).  For
the 188~Hz QPO in the 6--15.3~keV band we measure: $185^{+3}_{-3}$~Hz
(frequency), $18^{+17}_{-8}$~Hz (FWHM), $2.7^{+0.6}_{-0.4}$\% rms
(strength).  Assuming the frequency and width measured in the
6--60~keV band, we place 95\% confidence upper limits on the strength
of this QPO in the 2--6~keV band of 1.3\% rms, and in the 15.3--60~keV
band of 3.9\%.  The 268~Hz QPO is not significant in the 2--6~keV
band, and we place a 95\% confidence upper limit on its strength of
5.8\% rms.  For the 268~Hz QPO in the 6--15.3~keV band we measure:
$265^{+3}_{-3}$~Hz (frequency), $59^{+10}_{-8}$~Hz (FWHM), and
$6.5^{+0.4}_{-0.5}$\% rms (strength).  In the 15.3--60~keV band we
measure: $267^{+10}_{-14}$~Hz (frequency), $40^{+25}_{-17}$~Hz (FWHM),
and $6^{+1}_{-1}$\% rms (strength).  Particularly for the 188~Hz QPO,
the upper limits we obtain are not very constraining.

In order to test whether the 188~Hz and 268~Hz QPOs are present
simultaneously, we divided the 18 PCA observations into three equal
time segments and three even count rate slices, and analyzed these six
power density spectra.  In general, in full observations wherein we
clearly detect the 268~Hz QPO, the high-count-rate power density
spectra (and time segments with a relatively higher count rate)
suggest two QPOs consistent with the 188~Hz and 268~Hz QPOs we report.
In these segments, we do not find evidence for an alternating presence
of the two peaks as a function of time or count rate, or for a single
broad QPO feature which would suggest that a single QPO moves in
frequency between 188--268~Hz.  Individual features are usually not
significant in these slices due to limited statistics and upper limits
on the rms of features are not constraining.  In lower-count-rate
segments of full observations, and in segments of observations wherein
the 268~Hz QPO is not detected, if a feature is suggested it is
consistent with the 268~Hz QPO and not with a broader feature which
might indicate a single moving QPO.  We find no time segment or
count-rate slice in which a feature consistent with the 188~Hz QPO is
present individually, or stronger than a feature consistent with the
268~Hz QPO.  Although our investigation indicates the QPOs are more
likely simultaneous, it is nevertheless possible that we have observed
only one QPO feature which moves in frequency on short timescales.

\section{Discussion}
The 268~Hz QPO likely represents the first detection of a high
frequency QPO in outbursts of a transient BHC separated by a clear
quiescent period.  This QPO, and the 285~Hz QPO reported in the
1998--1999 outburst (Remillard et al.\ 1999a; Homan et al.\ 1999;
Homan et al.\ 2001) are very likely the same QPO, as the frequency,
FWHM, and strengths are very comparable.  Yet, in a number of ways the
1998--1999 and 2000 outbursts of \xtej~are significantly different.
In the former, a two-parameter model (e.g., $\dot{m}$ and an as-yet
unknown parameter) is required to describe the observed state
transitions, the high frequency QPOs are correlated with spectral
hardness, and the QPO parameters vary significantly with time.  The
2000 outburst of \xtej~is consistent with a FRED lightcurve with state
transitions accounted for only by variations in $\dot{m}$, the high
frequency QPO parameters show no clear correlation with spectral
hardness, and the QPO parameters vary little over time.  That the same
high frequency QPO is observed in such different outbursts, however,
likely indicates a fundamental similarity of the accretion flow
environments.

We do not detect the high frequency QPO in the initial LS.  This
finding is consistent with timing studies of the LS following the
outburst peak by Kalemci et al. (2001).  If the QPO frequency is
related to the Keplerian frequency at the innermost stable circular
orbit around the black hole, this may suggest that the accretion disk
is not filled entirely during the LS, and may recede slightly within
the VHS/IS.  Esin, McClintock, and Narayan (1997) describe an
advection-dominated accretion flow (ADAF) model for BHCs.  In this
model, the accretion disk cannot cool efficiently at low $\dot{m}$ and
the accretion flow becomes quasi-spherical close to the BH (the inner
disk radius is truncated at some distance from the marginally stable
circular orbit).  Thus, in this model, $\dot{m}$ governs whether the
inner region is hot and quasi-spherical, or an accretion disk.  With
the exception of the VHS, this model can describe the FRED lightcurve
of GS~1124$-$68 (Ebisawa et al.\ 1994).  As the 2000 outburst of
\xtej~also follows a FRED lightcurve, it might be well-suited to ADAF
modelling.  The similarity of accretion flow environments implied by
the detection of the same high frequency QPOs in the 1998--1999 and
2000 outbursts of \xtej~urges the modelling of non-FRED lightcurves
with the ADAF model as well.

In the 1998--1999 outburst of \xtej, QPOs reported at roughly 185~Hz
and 284~Hz (Remillard et al.\ 1999a; Homan et al.\ 1999) are likely
not distinct, but are more likely the same high frequency QPO: Homan
et al.\ 2001 note that within the hard states in the 1998--1999
outburst, the high frequency QPO may move between 102--284~Hz.  Our
investigation of the fast variability in the 2000 outburst of
\xtej~suggests that two high frequency QPOs may be present
simultaneously at approximately 188~Hz and 268~Hz. 

This evidence for simultaneous high frequency QPOs in \xtej~is similar
to the simultaneous high frequency QPOs in transient BHC/microquasar
GRO~J1655$-$40, at 450~Hz and 300~Hz (Strohmayer 2001a; the 300~Hz QPO
was discovered earlier by Remillard et al. 1999b).  In GRS~1915$+$105,
a QPO at 40~Hz has been discovered (Strohmayer 2001b), present
simultaneously with the previously-reported QPO at 67~Hz (Morgan,
Remillard, and Greiner 1997).  In all three systems, the QPO
frequencies are roughly in the ratio of 3:2.  In the case of
GRO~J1655$-$40, in the energy band where the 450 Hz QPO is strongest,
it is stronger than the 300~Hz QPO.  We do not find such a different
dependence of QPO amplitude on energy for the two peaks in \xtej~(see
Section 3).

The presence of two high frequency QPOs in BHCs, present
simultaneously, may be an emerging paradigm which will allow for new
constraints on models for QPO production.  If the QPOs observed in
GRO~J1655$-$40, GRS~1915$+$105, and \xtej~are analogous to the twin
kHz QPOs often observed in neutron star systems, then models for the
QPOs in neutron star systems which require reactions with a solid
stellar surface (e.g., the sonic point beat frequency model of Miller,
Lamb, \& Psaltis 1998) may be invalid.  Models for QPOs in neutron
star and BHC systems which instead rely upon General Relativistic
frequencies at the inner accretion disk edge (e.g., the relativistic
precession frequency model of Stella, Vietri, \& Morsink 1999) are not
excluded by the discovery of simultaneous high frequency QPOs in BHC
systems.

If the 450~Hz QPOs represent a modulation at the innermost stable
circular orbit of the accretion disk in GRO~J1655$-$40, which is
dynamically constrained to have a primary with mass of
$M~=~5.5-7.9~M_{\odot}$ (Shahbaz et al.\ 1999), this implies an
angular momentum parameter $j~=~cJ/GM^{2}$ of 0.15--0.5 (Strohmayer
2001).  Though certainly less concrete, the possibility of a black
hole with significant angular momentum parameter had previously been
inferred in spectral fits to data from GRO~J1655$-$40 (Zhang, Cui, \&
Chen 1997; Balucinska-Church and Church 2000), and BHC XTE~J1748$-$288
(Miller et al. 2001).  The mass of the primary in \xtej~was
constrained via optical work: $M_{1} > 7.4\pm0.7~M_{\odot}$ (Orosz et
al. 2001).  Very recent work has found that the most likely mass of
the compact object in \xtej~is 9.61~$M_{\odot}$, with a 3$\sigma$
range of $7.50~M_{\odot} \leq M_{1} \leq 13.18~M_{\odot}$ (Orosz et
al., private communication).  Although the most likely mass suggests a
black hole with non-zero $j$, the 3-sigma lower limit on the mass of
the compact object does not require a black hole with significant
angular momentum.  This suggests that the presence of simultaneous
high frequency QPOs may not be a unique signature of a black hole with
non-zero angular momentum.  The relativistic precession QPO model
(Stella et al.\ 1999) suggests that these black hole systems may all
have significant angular momentum.  We look forward to combining the
high-resolution spectroscopy available with \textit{Chandra} and
\textit{XMM-Newton} with the timing resolution of \textit{RXTE} to
further probe the question of black hole spin.

\section{Acknowledgements}
We wish to thank Jean Swank, Evan Smith, and the RXTE staff for
executing our Target of Opportunity program.  We are indebted to Jerry
Orosz for his generous communication of excellent work.  We wish to
acknowledge Ron Remillard for many useful discussions.  We thank the
anonymous referee for helpful suggestions.  R.\ W.\ was supported by
NASA through Chandra fellowship grants PF9-10010, which is operated by
the Smithsonian Astrophysical Observatory for NASA under contract
NAS8--39073.  C.\ K.\ acknowledges support from NASA through LTSA
grant NAG5--8496.  M.\ K.\ acknowledges support by the Netherlands
Organization for Scientific Research (NWO).  W.\ H.\ G.\ L.\
gratefully acknowledges support from NASA.  This research has made use
of the data and resources obtained through the HEASARC on-line
service, provided by NASA-GSFC.


\pagebreak

\begin{figure}
\figurenum{1}
\label{fig:Lightcurve}
\centerline{~\psfig{file=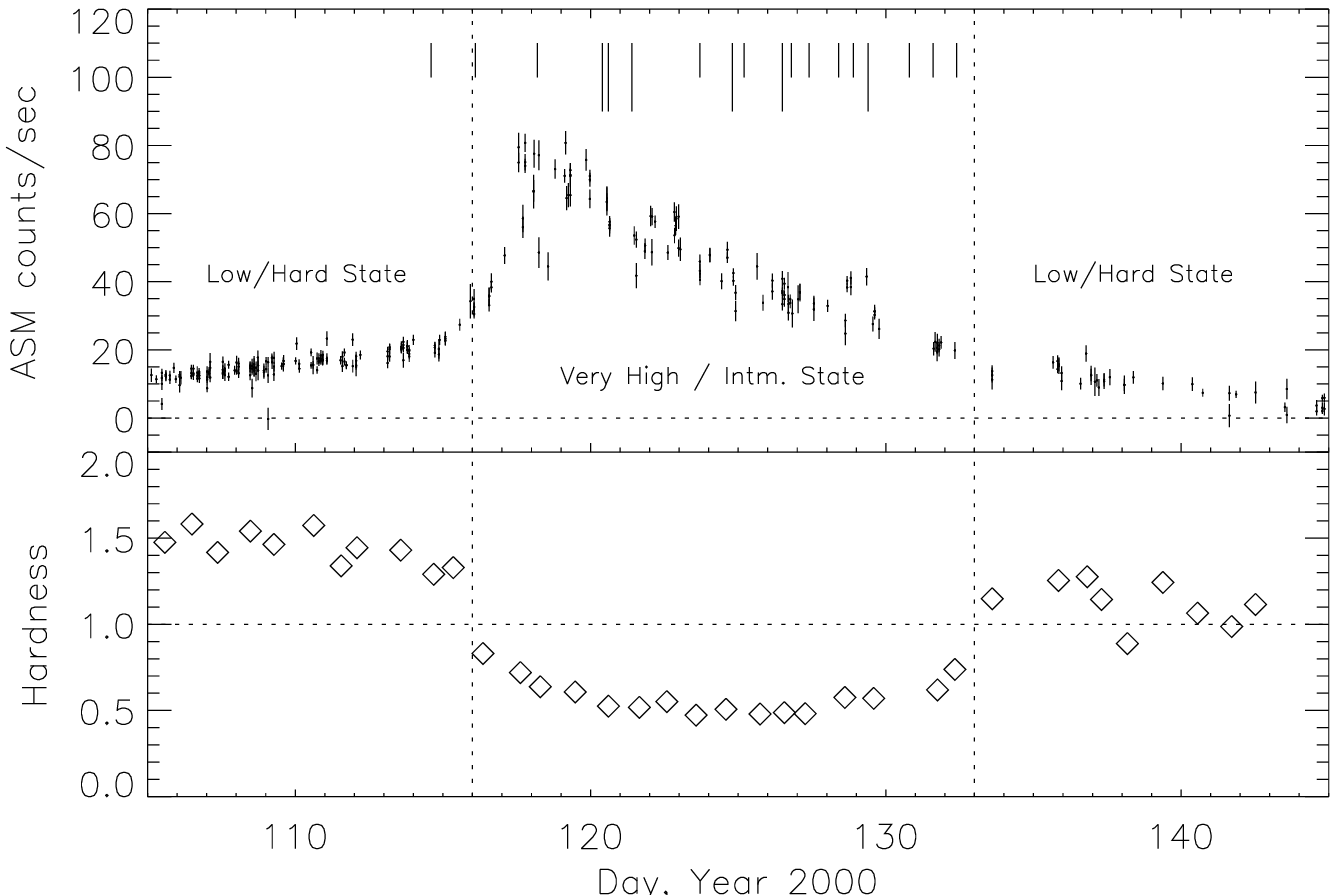,width=6.0in,height=4.0in}~}
\caption{The ASM dwell-by-dwell lightcurve with errors (1.5--12~keV),
and one-day averaged (5--12~keV)/(3--5~keV) hardness ratios, for the
April-May 2000 outburst of \xtej~(see also Miller et al. 2001b, Corbel
et al. 2001).  The source rise is typical of the LS.  It is followed
by a VHS/IS episode where the flux softens as the overall flux
increases rapidly, and finally decays in a LS into quiescence.  The
dashed vertical lines indicate the state transitions.  All of the high
frequency QPOs we report occur during the VHS/IS.  Vertical marks at
the top of the plot denote the days on which we observed
XTE~J1550$-$564; longer marks denote observations in which a high
frequency QPO is found.}
\end{figure}

\begin{table}[t]
\caption{High Frequency QPOs in \xtej}
\begin{small}
\begin{center}
\begin{tabular}{llllllll}
Day & Frequency & FWHM & rms & rms & rms & rms & rms \\ 
~  & ~ & ~    & (6-60 keV) & (2-60 keV) & (2-6 keV) & (6-15.3 keV) & (15.3-60 keV) \\
(2000) & (Hz) & (Hz) & (\%) & (\%) & (\%) & (\%) & (\%) \\
\tableline
\tableline
120.4$\dag$ & 276(4) & $61^{+17}_{-12}$ & 7.4(6) & 2.6(2) & 2.0(2) & 7.0(5) & 4(4)\\
~ & ~ & ~ & [6.2$\sigma$] & ~ & ~ & ~ & ~ \\
\tableline
120.6$\dag$ & $275^{+4}_{-6}$ & $40^{+25}_{-16}$ & $6.9^{+0.8}_{-0.6}$ & 1.9(2) & 1.5(2) & 6.5(5) & 7(2) \\
~ & ~ & ~ & [5.8$\sigma$] & ~ & ~ & ~ & ~ \\
\tableline
121.4$\dag$ & 265(7) & $65^{+26}_{-16}$ & $7.7^{+1.0}_{-0.8}$ & 3.2(2) & 2.7(3) & 7.4(7) & 8(4) \\
~ & ~ & ~ & [4.8$\sigma$] & ~ & ~ & ~ & ~ \\
\tableline
124.8 & 261(8) & $53^{+18}_{-13}$ & $5.3^{+0.7}_{-0.6}$ & 2.1(3) & 1.8(5) & 5.0(6) & 6(5) \\
~ & ~ & ~ & [4.4$\sigma$] & ~ & ~ & ~ & ~ \\
\tableline
126.5 & 249(6) & $58^{+24}_{-17}$ & $5.6^{+0.7}_{-0.6}$ & 2.6(2) & 1.9(4) & 5.4(4) & 7(3) \\
~ & ~ & ~ & [4.7$\sigma$] & ~ & ~ & ~ & ~ \\
\tableline
129.4 & 251(6) & $45^{+20}_{-13}$ & $5.4^{+0.8}_{-0.6}$ & 2.0(4) & $<$2.7 & 4.7(6) & 12(2) \\
~ & ~ & ~ & [4.5$\sigma$] & ~ & ~ & ~ & ~ \\
\tableline
\tableline
\end{tabular} ~\vspace*{\baselineskip}~\\ \end{center} 
\tablecomments{QPO parameters for those observations wherein a high
frequency QPO is detected.  Detailed above are the day of the year
2000 on which the observation was made, and the Frequency, FWHM, and
rms amplitude of each QPO, in 5 energy bands.  The rms values noted in
other bands are measured by fixing the frequency and FWHM to that
measured in the band where the QPO is most significant.  Errors are
1$\sigma$ confidence intervals, customary for timing studies.  $\dag$
denotes observations made in an instrumental mode that samples the
full PCA energy bandpass at a slightly different binning.  The bands
for these observations are (in the order that appears left-to-right
above): 8-57 keV, 2-57 keV, 2-8 keV, 8-15.3 keV, and 15.3-57 keV.
95\% confidence upper limits (6--60~keV) on the strength of a high
frequency QPO in observations made on days 114.6, 116.1, 118.2, and
123.7 range between 2.7-5.2\% rms.  We find upper limits in the range
of 5.1--6.4\% rms for a high frequency QPO in observations made on
days 125.2, 126.8, 127.4, 128.4, 128.9, 130.8, 131.6, and 132.4 (any
feature would be significant at less than 2$\sigma$).}
\vspace{-1.0\baselineskip}
\end{small}
\end{table}

\pagebreak

\begin{figure}
\figurenum{2}
\label{fig:Parameters}
\centerline{~\psfig{file=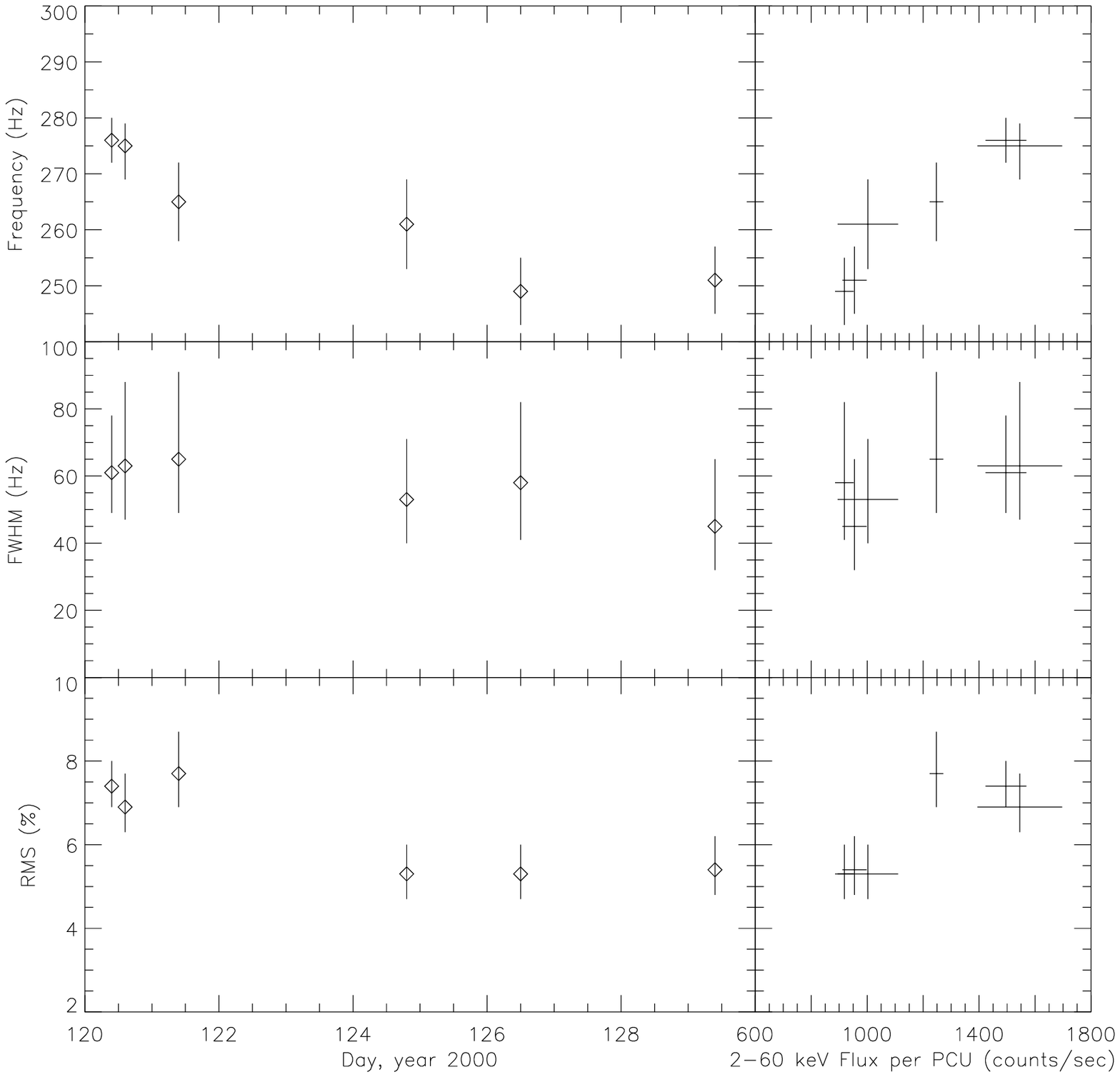,width=6.0in,height=6.0in}~}
\caption{QPO Frequency, FWHM, and rms plotted versus time and flux.
Displayed errors in QPO parameters and flux are 1$\sigma$ confidence
errors.}
\end{figure}

\pagebreak

\begin{figure}
\figurenum{3}
\label{fig:Hardness}
\centerline{~\psfig{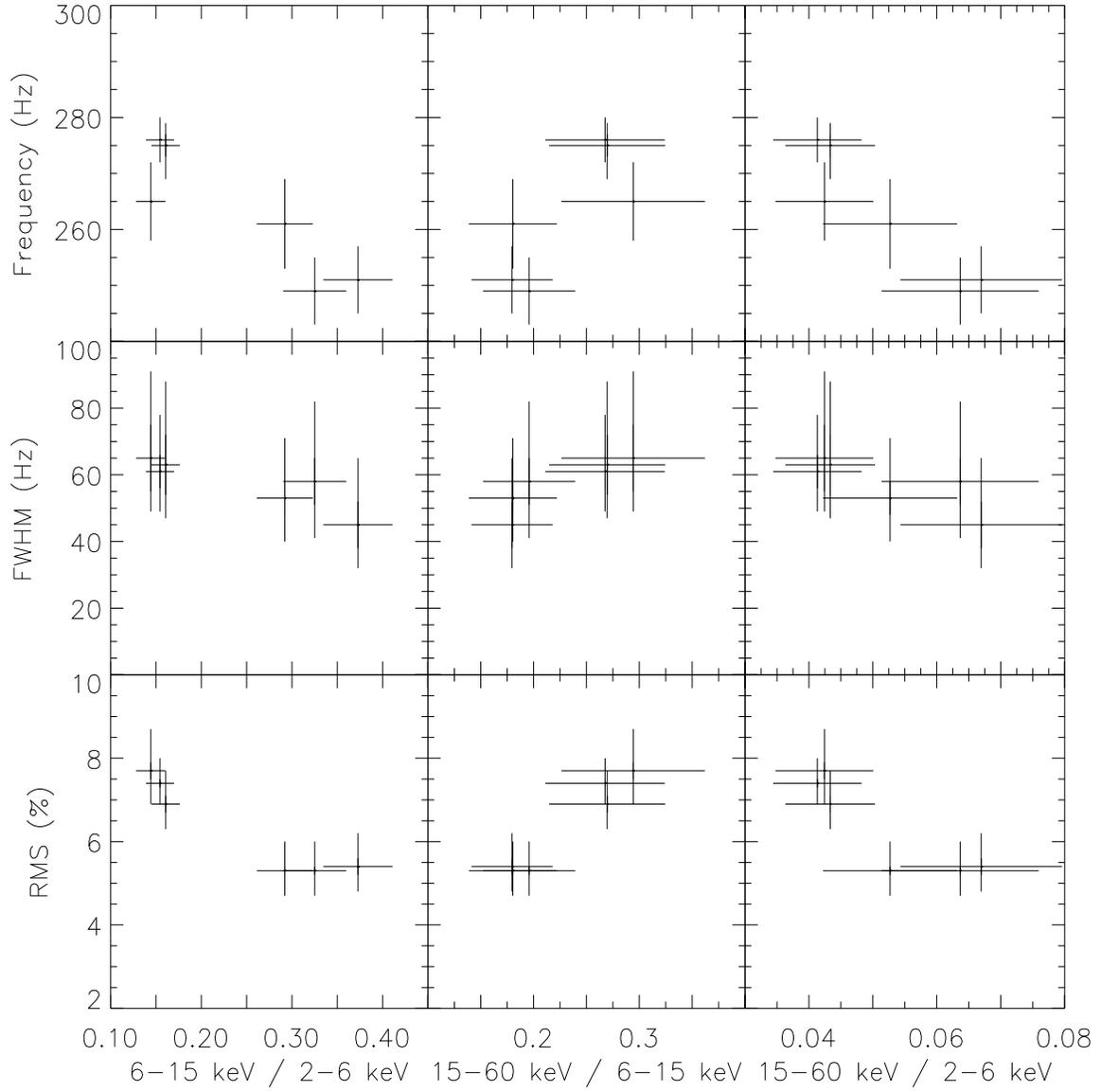}~}
\caption{QPO Frequency, FWHM, and rms plotted versus three hardness
ratios.  Displayed errors in QPO parameters and hardness ratios are
1$\sigma$ confidence errors.}
\end{figure}

\pagebreak

\begin{figure}
\figurenum{4}
\label{fig:Peaks}
\centerline{~\psfig{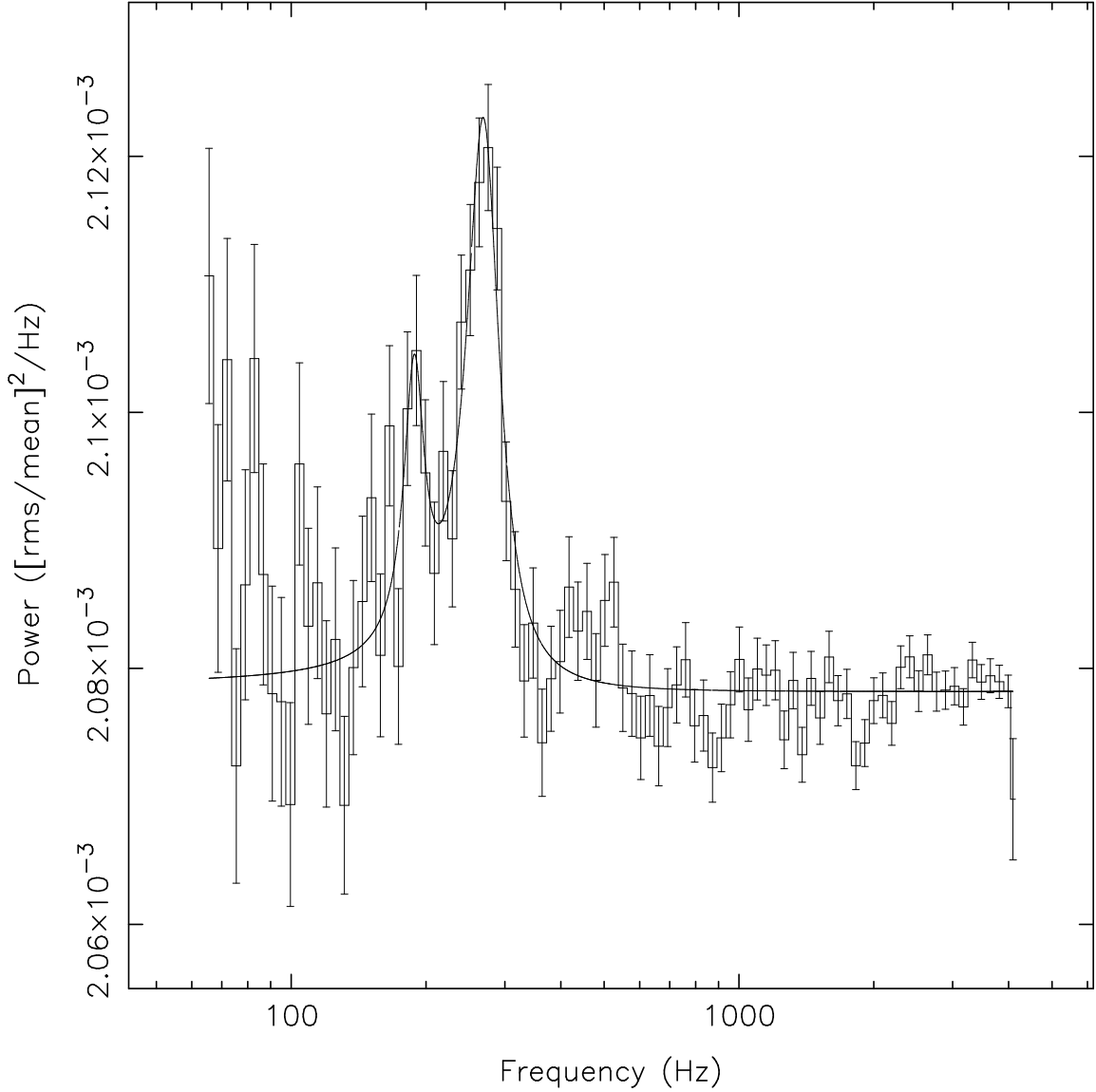}~}
\caption{All observations with high frequency QPOs added together
(6-60 keV, centroid frequencies not shifted).  The higher-frequency
QPO: 268$\pm$3~Hz, $56^{+8}_{-7}$~Hz FWHM, $6.2^{+0.4}_{-0.4}$\% rms,
significant at 7.8$\sigma$.  The lower-frequency QPO: 188$\pm$3~Hz,
$24^{+33}_{-9}$~Hz FWHM, $2.8^{+0.9}_{-0.4}$\% rms, significant at
3.5$\sigma$.  Quoted errors in QPO parameters are 1$\sigma$ confidence
errors, and significances are single-trial (here, there are 22 trials
for the 268~Hz QPO, and 48 trials for the 188~Hz QPO).}
\end{figure}

\end{document}